# An Optimization Framework For Online Ride-sharing Markets

Yongzheng Jia[1], Wei Xu[1] Xue Liu[2]

[1]Institute of Interdisciplinary Information Sciences, Tsinghua University

[2]School of Computer Science, McGill University

jiayz13@mails.tsinghua.edu.cn, weixu@mail.tsinghua.edu.cn, xueliu@cs.mcgill.ca

*Abstract*—Taxi services and product delivery services are instrumental for our modern society. Thanks to the emergence of sharing economy, ride-sharing services such as Uber, Didi, Lyft and Google's Waze Rider are becoming more ubiquitous and grow into an integral part of our everyday lives. However, the efficiency of these services are severely limited by the sub-optimal and imbalanced matching between the supply and demand. We need a generalized framework and corresponding efficient algorithms to address the efficient matching, and hence optimize the performance of these markets. Existing studies for taxi and delivery services are only applicable in scenarios of the one-sided market. In contrast, this work investigates a highly generalized model for the taxi and delivery services in the market economy (abbreviated as "taxi and delivery market") that can be widely used in two-sided markets. Further, we present efficient online and offline algorithms for different applications. We verify our algorithm with theoretical analysis and trace-driven simulations under realistic settings.

## I. INTRODUCTION

Recent years witness the rapid development of the online ride-sharing applications (such as Uber, Didi, Lyft and Google's Waze Rider), online e-commerce shopping and the corresponding product delivery services (such as Google Express and Amazon Prime Now). For these services to be practical, efficient task-scheduling algorithms design has become an ever-more important research area. This new area of research is important for the emergence of the sharing economy and has attracted wide attentions from researchers in the networking community, design and scheduling community, as well as researchers from operational research and economics. These services exist in the form of *two-sided market*, also known as the *matching market* [1]. In a two-sided market, there are two distinct user groups, the *workers* and the *customers*. The services need to provide each group of users with benefits. For example, in the ride-sharing market, the services (such as Uber) aim to reduce ride cost for the customers (i.e. riders) and at the same time, generate sizable income for the workers (i.e. drivers).

Let us take a closer look at the online ride-sharing applications. There are a large number of drivers and riders (a.k.a passengers) in the market. Each task (i.e. dispatching) can be described as a matching between the willing driver and the rider that takes the rider from one place (i.e. source) to another place (i.e. destination) during a specific period. When a rider places an order in the application (e.g. the Uber App), the system must notify the drivers and give instant response to the rider. The response includes whether the order can be served, and if so, the platform chooses a willing driver to serve the task (i.e. order), or the order fails. An efficient application should benefit both the drivers and the riders. For on-demand product delivery services, customers place orders online, and the service providers deliver the products/goods to the customers within the promised time frame, e.g. overnight, 2-day or 7-day. The platform needs to efficiently match the demand from customers and the supply from workers (drivers). It also needs to provide the drivers with determined travel/delivery plans to fulfill the orders in time.

In the above example scenarios, we see that it is crucial to schedule the tasks to the workers efficiently based on the supply and demand information in these two-sided markets. In this work, we first discuss that these two-sided markets can be formulated as a generalized mathematical model. We propose a theoretical two-sided market model that has two dynamic user groups: *workers* (drivers) and *customers*. Each order from customers has a *source* and *destination*, and has corresponding *start time* and *end time*; each worker has her working schedule with her start time and finish time every day; the platform calculates the cost and payoff of each order. Further in our model, we also consider the specific surge pricing model similar to that used by Uber [2]. In our model, we consider two objectives: One is maximizing the total profits of the drivers, and the other is maximizing the social welfare. The total profits of the drivers are the producer surplus without considering the utility to the customers, which is the total benefits achieved by the drivers. The social welfare is the sum of the producer surplus and the consumer surplus (i.e. the total utility to the customers), which is the metrics of the total benefits achieved by the two groups of users in the market.

Solving the optimization problems for such dynamic two-sided markets is not easy. There are two fundamental challenges to overcome. *First*, the algorithms need to be efficient enough to deal with a large number of workers and customers in the market. The ride-sharing market is huge, and it is essential that we can partition the problem. Therefore the algorithms have to be distributed. In real scenarios, we can partition the map in city's scale, and then design algorithms to deal with the tasks in each city. However, for some big cities such as NYC, London, and Beijing, there are often millions of passengers and tens of thousands of drivers available every day. In practice, it is not practical to further partition the problem into smaller divisions (i.e. districts). This is because the riders and drivers generally travel across the city. Hence the service should cover the entire city. *Second*, although offline algorithms can work for traditional e-commerce and product delivery services. Ride-sharing services and instant product delivery services need the platform to give instant responses (i.e. real-time responses) to the customers, making online algorithms necessary.

Existing efforts for taxi and delivery markets mostly focus on the models from the *Vehicle Routing Problems (VRP)*. However, most of the work can only deal with specific offline scenarios and cannot scale to larger models with a large amount of data in a distributed way. Further, most of these models are too detailed and thus limited to specific one-sided market. In contrast, this work investigates a more generalized model for the two-sided market. The model applies in both offline and online cases. The contributions of this paper are summarized as follows.

*First*, we establish a generalized model for the Internet taxi and delivery markets. The model can efficiently address the inefficient matching between the supply and demand in these two-sided markets even with surge pricing mechanism. Our model can be used in both centralized and distributed scenarios. We construct the task map to clearly show the relationship between the workers and the customers in the markets, and formulate the optimization problem with overall social welfare as the objective function.

*Second*, we propose an approximation algorithm to solve the above problem in an offline setting (i.e. we have all the travel plans in advance). We transfer the offline problem to the *multiple disjoint paths* (MDP) problem [3], and the goal is to find a set of node-disjoint paths in a directed acyclic graph with maximum total values. We carefully design a greedy algorithm and prove that the algorithm has a tight approximation ratio which applies to the ride-sharing market very well.

*Third*, we propose two heuristic algorithms that can be used for both offline and online settings. We show that our algorithms have good performance ratio by comparing our algorithms with the theoretical upper bound using real-trace simulations. We also show the effectiveness to apply our algorithms into real markets from the simulation results.

The rest of the paper is organized as follows. We discuss related work in Section II and present the problem model in Section III. The deterministic approximate algorithm for offline cases is given in Section IV and the online algorithms are given in Section V. We present trace-driven evaluation results in Section VI and conclude the paper in Section VII.

## II. RELATED WORK

Research interests on online "sharing" and "gig" economy platforms have increased significantly these years. [4] provides a broad discussion of sharing economics and two-sided markets. The world's largest ride-sharing company, Uber, has the mechanism that dynamically prices trips using a system known as "surge pricing". [2] gives a detailed measure of how the dynamic pricing of tasks influences the supply of labor in the market. [5] observes the Uber's surge pricing mechanism by the datasets generated from the Uber mobile app to tackle the questions about fairness and transparency of the system. Based on the surge pricing mechanism, [6] shows that ride-sharing services not only dramatically increase the usage of drivers and their cars, but also cut costs to the riders. [7] develops a modeling framework for studying a decentralized equilibrium based market study, and optimizes the fleet size and pricing policy for a given urban area.

To the best of our knowledge, our work is the first to study the sharing economics of taxi and delivery markets with both online and offline models and their corresponding solutions. Existing efforts focus on the one-sided market under the settings of Vehicle Routing Problem (VRP). The VRP with time windows [8] has some similarities with our model in terms that each node has a deadline and a release-time. The goal is to visit as many nodes as possible within the "time-windows". [9] gives a deterministic algorithm with $O(log^2 N)$ approximation for the VRP with time-windows where $N$ is the number of nodes in the graph. People have studied various heuristics [10] [11] [12] such as local search, simulated annealing and genetic algorithms, as well as branch and bound methods [13] [14] to solve this kind of VRP as well.

Our deterministic algorithm follows the formulation of the MDP problem. The maximum *edge-disjoint* paths problem in directed graphs is formulated in [15] and [3]. [15] gives a greedy algorithm with $O(\sqrt{m})$ approximation ratio using a multi-commodity-flow-based LP relaxation. In our approximation algorithm solution, we will use another approximation ratio instead of $O(\sqrt{m})$. We show why our approximation ratio is better suited for the ride-sharing market and prove our bound is tight in Section IV.

## III. PROBLEM MODEL

In this section, we propose the model of the two-sided market. There are two groups of users. We use the taxi and delivery markets (referred to as "the market") as examples throughout the paper. We use *drivers* to represent the users who provide taxi or delivery services, and *customers* to represent the users who receive the services. *Tasks* represent the taxi and delivery services ordered by the customers.

### A. Market Configuration

There are total $N$ drivers available in the market. We consider the generalized model that each driver reveals her travel plan with a given source and a destination everyday before she starts working (e.g. they can be the driver's home address). For each driver $n \in [N]$, she starts her travel plan from the source location $s_n$ at time $t_n^-$ to her destination location $d_n$ at time $t_n^+$ that $t_n^- < t_n^+$. We depict the geo-location information of driver $n$'s source and destination using two tuples $s_n = (u_n^-, v_n^-)$ and $d_n = (u_n^+, v_n^+)$ where $u_n$ and $v_n$ are the latitude and the longitude respectively. We use $[X] = \{1, 2, \ldots, X\}$ to denote the set of $X$ elements throughout the paper, e.g., $[N] = \{1, 2, \cdots, N\}$ is the set of drivers.

There are a total of $M$ tasks submitted by the customers in the market during a certain time period. Each task has a publishing time $\bar{t}_m$ that the customer submits the task to the market. Being the same as the setting of the drivers, each task also has its source and destination locations such that task $m \in [M]$ starts at source $\bar{s}_m = (\bar{u}_m^-, \bar{v}_m^-)$ at time $\bar{t}_m^-$ and ends at its destination $\bar{d}_m = (\bar{u}_m^+, \bar{v}_m^+)$ at time $\bar{t}_m^+$. Note that $\bar{t}_m < \bar{t}_m^- < \bar{t}_m^+, \forall m \in [M]$, since task $m$ should be submitted before it starts. In the online scenarios, as we do not know the accurate start time and end time of each task in advance, we can use $\bar{t}_m^-$ as the deadline of the start time of task $m$, and $\bar{t}_m^+$ as the deadline of the end time of task $m$. Task $m$ may start earlier than $\bar{t}_m^-$ and finish earlier than $\bar{t}_m^+$. We use $b_m$ to denote the customer's valuation for task $m$, which is her WTP (i.e. willingness to pay) for task $m$. The customer will only admit to publish the task when her WTP is higher than the price of the task. Otherwise, she will gain negative utility and hence refuse to publish her task.

Each task $m \in [M]$ also has a certain price $p_m$ calculated by the platform as the payoff to the driver. Note that the task will only be published when $p_m \leq b_m$. Unlike traditional taxi services, platforms like Uber dynamically adjust their prices using a *surge pricing mechanism* [2]. The price rate, also named as the *Surge Multiplier* (SM), increases when demand is higher than supply for a given geographic area. Customers are informed of the higher fare before requesting the service. Uber drivers are also aware of the surge pricing when the orders are published. However, no matter what pricing mechanism the platform adopts, the system calculates the price of the task and publishes to both its customers and drivers, therefore price $p_m$ can be treated as a constant attribute of a given task in the market setting.

## B. Task Map Construction

To demonstrate the relationship between the drivers and tasks in the market, we construct the task map for each driver. We model the task map with a directed acyclic graph (DAG) to illustrate whether driver $n \in [N]$ can take task $m' \in [M]$ in time after finishing task $m \in [M]$. Each task is a node in the graph, an arc (directed edge) in the graph implies the availability for the driver to take another task after finishing her previous one. For each driver's task map, there are two special nodes: the source and the destination, which are labeled by number 0 and $-1$ respectively. Therefore the node set of driver's task map is $[\hat{M}] = \{-1, 0\} \cup [M]$. The task map is dynamic in the online scenarios, and we initialize the finish time of task $m$ by using $\bar{t}_m^+$. When the task $m$ finishes before $\bar{t}_m^+$, we use the real finish time. We show how to deal with the real finish time for the online scenarios in Section V

To estimate the travel time from one place to another in the graph, we first estimate the travel distances. We need to consider two types of distances, one is the travel distance from the destination of one task to the source of the next one, and the other is the travel distance from the source to the destination of the same task. We denote the estimation of the travel distance for driver $n$ to travel from the destination of task $m$ to the source of the next task $m'$ (i.e. driving empty) as $d_{n,m,m'}$, and the estimation of the travel distance for driver $n$ to travel from the source to the destination of the same task $m$ (i.e. driving the customer) is $\hat{d}_{n,m}$.

Now we can estimate the travel time from one node to another using an estimated driving speed of the driver. Let $l_{n,m,m'}$ be the estimated travel time from the destination of task $m$ to the source of task $m'$ for driver $n$ (i.e. driving empty), and let $\hat{l}_{n,m}$ be the estimated travel time for driver $n$ to travel from the source to the destination of the same task $m$ (i.e. driving customers). Let $c_{n,m,m'}$ be the estimated travel cost from the destination of task $m$ to the source of the next task $m'$ for driver $n$, and $\hat{c}_{n,m}$ be the travel cost for driver $n$ to travel from the source to the destination of the same task $m$.

To construct the task map, We use indicator $h_{n,m,m'} \in \{0, 1\}, \forall n \in [N], m, m' \in [\hat{M}]$ to denote whether there is an arc from node $m$ to $m'$ in driver $n$'s task map. Figure 1 shows an example task map of driver $n \in [N]$. The source of driver $n$ is labeled by number 0, the destination of driver $n$ is labeled by $-1$ and the tasks are labeled by integers $\{1, 2, \cdots, M\}$. The prerequisite of driver $n$ to take task $m$ is that he has enough time to travel from the source to the destination of task $m$. Let indicator variable $\hat{h}_{n,m}$ denotes whether driver $n$ can take task $m$, with $\hat{h}_{n,m} = 1$ indicating a "yes" as follows:

$$\hat{h}_{n,m} = 1 \Leftrightarrow (\hat{l}_{n,m} \leq \bar{t}_m^+ - t_m^-), \quad \forall n \in [N], m \in [M]. \quad (1)$$

For the arcs from the source (labeled 0) to any task $m$,

$$\begin{aligned} h_{n,0,m} = 1 \Leftrightarrow &\hat{h}_{n,m} \wedge (l_{n,0,m} \leq \bar{t}_m^- - t_n^-) \\ &\wedge (l_{n,m,-1} \leq t_n^+ - \bar{t}_m^+), \quad \forall n \in [N], m \in [M]. \end{aligned} \quad (2)$$

As (2) shows, if there is an arc from the source of driver $n$ to task $m$, we need driver $n$ to have enough time to travel from her source location to the source of task $m$, and also have enough time to travel from the destination of task $m$ to driver $n$'s destination. If $h_{n,0,m} = 1$, we also set $h_{n,m,-1} = 1$, and then draw one arc from node 0 to $m$, as well as another arc from node $m$ to $-1$ on driver $n$'s task map.

For the arc from one node of task $m$ to the next task $m'$, driver $n$ should have enough time to travel from the destination of task $m$ to the source of task $m'$.

$$\begin{aligned} h_{n,m,m'} = 1 \Leftrightarrow &\hat{h}_{n,m} \wedge \hat{h}_{n,m'} \wedge (l_{n,m',-1} \leq t_n^+ - \bar{t}_{m'}^+) \\ &\wedge (l_{n,m,m'} \leq \bar{t}_{m'}^- - \bar{t}_m^+), \forall n \in [N], m \in [M], m' \in [M]. \end{aligned} \quad (3)$$

If $h_{n,m,m'} = 1$ then also set $h_{n,m',-1} = 1$, there is an arc from $m$ to $m'$ and another arc from $m'$ to $-1$.

It will take $(M^2 + 2M)$ iterations to calculate all the values of $h_{n,m,m'}$ for driver $n$. Therefore the time complexity to construct the task map of all the $N$ drivers is $O(NM^2)$.

To clearly show the tasks for each driver, we investigate the task list of driver $n$ during her work using graph theory.

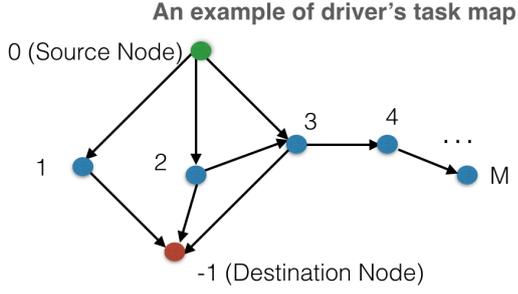

Fig. 1. shows a simple example task map of driver $n$. The driver can take one task among task 1, task 2 and task 3. She can also take two tasks, and that is to take task 3 after finishing task 2.

Driver $n$ takes a sequence of tasks, that can be illustrated as a path in driver $n$'s task map, from her source node 0, then to the task nodes in her task list one by one, and finally to her destination node $-1$. Therefore the path is a flow from the source to the destination, such that in/out degrees of each task node are both 1 or both 0 (if the task is not taken), the source node has out-degree 1 and in-degree 0, and the destination node has out-degree 0 and in-degree 1. Then we can clearly formulate our optimization problem by applying the network flow model.

We summarize the key notations throughout this paper in TABLE I as follows.

TABLE I
KEY NOTATIONS

| | | | |
|---|---|---|---|
| $[N]$ | set of drivers $\{1, 2, \cdots, I\}$ | $N$ | # of drivers |
| $[M]$ | set of tasks $\{1, 2, \cdots, M\}$ | $M$ | # of tasks |
| $[\hat{M}]$ | nodes $\{-1, 0, 1, 2, \cdots, M\}$ in task map | | |
| $t_n^-, t_n^+$ | start / end time of driver $n$ | | |
| $s_n, d_n$ | source / destination of driver $n$ | | |
| $\bar{t}_m^-, \bar{t}_m^+$ | estimated start / end time of task $m$ | | |
| $\bar{s}_m, \bar{d}_m$ | source / destination of task $m$ | | |
| $p_m$ | payoff for task $m$ | | |
| $b_m$ | customer's WTP for task $m$ | | |
| $c_{n,m,m'}$ | travel cost from destination of $m$ to source of $m'$ for driver $n$ | | |
| $\hat{c}_{n,m}$ | travel cost from source of $m$ to destination of $m$ for driver $n$ | | |
| $h_{n,m,m'}$ | indicator of whether there is an arc from node $m$ to node $m'$ in driver $n$'s task map | | |
| $x_{n,m}$ | whether task $m$ is assigned to driver $n$ | | |
| $y_{n,m,m'}$ | whether driver $n$ continues to take task $m'$ after finishing task $m$ | | |

Now we formulate our optimization problem. We have two objectives; we will first give a drivers' profit maximization formulation, and then give a social welfare maximization formulation.

### C. Drivers' Profit Maximization

We first propose a drivers' profit maximization framework to address the inefficient matching of the drivers and customers in the taxi and delivery markets, with the constraints to guarantee the feasibility to arrange a sequence of tasks to a certain driver by applying the network flow model. The decision variable $x_{n,m}$ indicates whether task $m$ is assigned to driver $n$ in the market. We also involve decision variable $y_{n,m,m'}$ at a micro level, denoting whether driver $n$ takes task $m'$ after finishing task $m$. In the network flow model, variable $y_{n,m,m'}$ is the actual flow among the nodes in the network graph. The drivers' profit maximization framework is formulated as follows:

$$Z : \text{maximize} \sum_{n \in [N]} \sum_{m \in [M]} x_{n,m} p_m - \Big( \sum_{n \in [N]} \sum_{m \in [M]} x_{n,m} \hat{c}_{n,m}$$
$$+ \sum_{n \in [N]} \sum_{m \in [\hat{M}]} \sum_{m' \in [\hat{M}]} y_{n,m,m'} h_{n,m,m'} c_{n,m,m'} - \sum_{n \in [N]} c_{n,0,-1} \Big).$$
(4)

s.t.
$$\sum_{n \in [N]} x_{n,m} \leq 1, \quad \forall m \in [M]; \tag{5a}$$

$$\sum_{m \in [M]} x_{n,m} p_m \geq \sum_{m \in [\hat{M}]} \sum_{m' \in [\hat{M}]} y_{n,m,m'} h_{n,m,m'} c_{n,m,m'}$$
$$+ \sum_{m \in [M]} x_{n,m} - c_{n,0,-1}, \quad \forall n \in [N]; \tag{5b}$$

$$\sum_{m' \in [\hat{M}]} h_{n,0,m'} y_{n,0,m'} = 1, \quad \forall n \in [N]; \tag{5c}$$

$$\sum_{m \in [\hat{M}]} h_{n,m,-1} y_{n,m,-1} = 1, \quad \forall n \in [N]; \tag{5d}$$

$$\sum_{m \in [\hat{M}]} h_{n,m,m'} y_{n,m,m'} = x_{n,m'}, \forall n \in [N], m' \in [M]; \tag{5e}$$

$$\sum_{m' \in [\hat{M}]} h_{n,m,m'} y_{n,m,m'} = x_{n,m}, \forall n \in [N], m \in [M]; \tag{5f}$$

$$x_{n,m} \in \{0, 1\}, \quad \forall n \in [N], m \in [M]; \tag{5g}$$

$$y_{n,m,m'} \in \{0, 1\}, \quad \forall n \in [N], m \in [\hat{M}], m' \in [\hat{M}]. \tag{5h}$$

The objective function (4) is the total revenue of all the tasks assigned to the drivers in the market subtracts the *excess cost* (not the total cost) of all the drivers to complete these tasks. The first term is the total revenue in the market. The second term is the total travel cost for the drivers to finish the tasks from the source to the destination of the same task. The third term is the total travel cost of all the drivers from the current task destination to the source of the next task, as well as the cost from the driver's source to the first task and the cost from the last task to the driver's destination. The forth term is the original travel cost for the drivers from source to destination without taking any tasks. Therefore the formula in the parentheses is the *excess cost* for all the drivers to complete their tasks.

For the constrains of the optimization problem, (5a) means that each task is assigned to at most one driver. (5b) is the *individual rationality* constraint that for each driver, whose total revenue must be no less than the *excess cost* to finish her tasks. (5c)-(5f) are flow conservation constrains in the network flow model. (5c) means the out-degree for the source of each driver is 1, (5d) means the in-degree for the destination of each driver is 1. (5e) and (5f) together show that if task $m$ is assigned to driver $n$, the in/out degrees of task node $m$ are 1, and otherwise, the in/out degrees of task node $m$ are 0. (5g) and (5h) show that all the decision variables of $x_{n,m}$ and $y_{n,m,m'}$ are binary variables in $\{0,1\}$.

### D. Social Welfare Maximization

We now propose a social welfare maximization formulation by taking the utility to the customers into consideration. The social welfare is the sum of the drivers' total profits and the utility to the customers, and the goal is to benefit both the drivers and the customers in the market.

We formulate the objective function of social welfare maximization in (6). Note that the only difference between (6) and (4) is switching $p_m$ into $b_m$, since the price is offset when considering the utility to the customers. For the constraints, we add (7a) since the customers will never gain negative utility by *individual rationality*. Actually, if $b_m < p_m$ for some $m$, the task may not even be published, such that we can ignore constraint (7a) in real scenarios.

$$\hat{Z} : \text{maximize} \sum_{n \in [N]} \sum_{m \in [M]} x_{n,m} b_m - \Big( \sum_{n \in [N]} \sum_{m \in [M]} x_{n,m} \hat{c}_{n,m} + \sum_{n \in [N]} \sum_{m \in [\hat{M}]} \sum_{m' \in [\hat{M}]} y_{n,m,m'} h_{n,m,m'} c_{n,m,m'} - \sum_{n \in [N]} c_{n,0,-1} \Big).$$
(6)

s.t.

$(5a) - (5h);$

$$\sum_{n \in [N]} x_{n,m}(b_m - p_m) \geq 0, \forall m \in [M].$$
(7a)

### E. Problem-solving ideas

In the real markets, it is hard to formulate the social welfare, since it is always hard to accurately estimate a certain customer's WTP for a ride. Actually, optimizing the drivers' total profits is enough to improve the efficiency of the ride-sharing markets. Therefore in the following sections, we will focus on solving the optimization problem of the drivers' total profits. We also declare that we can use the same algorithms given in Section IV and Section V to solve the social welfare maximization problem.

Our problem (4) can be reformulated to a special case of the MDP problem in Section IV, which is shown to be NP-hard in [16]. However, solving the relaxed problem can be done in polynomial time [17]. The relaxed problem is formulated as follows:

$$Z_f : maximize \quad (4).$$

s.t.

$(5a) - (5f);$

$$x_{n,m} \in \{0,1\}, \quad \forall n \in [N], m \in [M];$$
(8a)

$$y_{n,m,m'} \in \{0,1\}, \quad \forall n \in [N], m \in [\hat{M}], m' \in [\hat{M}].$$
(8b)

Let $Z^*$ denote the best integral solution of $Z$ in (4), and let notation $Z_f^*$ denote the best fractional solution of the relaxed problem. We have $Z_f^* \geq Z^* = OPT$, such that $Z_f^*$ provides an upper bound of the optimization solution $Z^*$.

## IV. A Deterministic Offline Solution

In this section, we propose a greedy algorithm to solve the optimization problem $Z$ in (4) by adjusting our problem to the MDP (i.e. maximum node-disjoint paths) model, then give the theoretical analysis of the algorithm. We further illustrate why our approximation ratio is better suited for the ride-sharing market.

### A. Transfer to the MDP problem

From graph theory, our problem can be categorized to the family of MDP problems. People have proposed algorithms to find a set of edge-disjoint paths (EDP) in a DAG to maximize the total value of the paths. [3] gives the formulation and an approximation algorithm solution of the EDP problems. Our problem is similar to the EDP, the difference is that we aim to find *weighted node-disjoint paths* (instead of edge-disjoint paths) with max total value. In the configurations discussed in the previous section, we give a multi-commodity-flow network using a DAG with a set of source-destination pairs representing the drivers. We model a driver's task list by a path from her source node to her destination node. Our objective is to find the set of paths with max total value in the network such that no two paths intersect at the same node, since each task can only be served by at most one driver. To sum up, each source-destination pair represents a driver and each path in the DAG is a possible task list for a driver.

Here we give another integer programming formulation of our problem by using the node-disjoint paths model. We use an exponential-sized paths formulation for a better demonstration of the problem nature, while in practice we do not need the exhaustion of all the possible paths in our solution. The two formulations of (4) and (9) are equivalent. We merge all the task maps of the $N$ drivers into a big graph $G$, and therefore $G$ is a DAG that contains all the source nodes, destination nodes, and task nodes. Let $\mathcal{P}_i$ denote all the paths in the graph $G$ from $s_i$ to $d_i$ for driver $i$. For each path $\pi$ in $\cup_i \mathcal{P}_i$, we have a variable $r_\pi$ for the profit of the path and a binary variable $f_\pi$ of whether path $\pi$ is selected in the solution. $r_\pi$ can be calculated by the summation of the total value of the tasks subtracting the excess cost (defined in Eq. (4)) of the path. For each driver $i$, we use binary variable $x_i$ to indicate whether each driver has 0 (if no task for the driver) or 1 task list.

$$Z : \text{maximize} \sum_{\pi \in \cup_i P_i} f_\pi r_\pi.$$
(9)

s.t.

$$\sum_{\pi \in \mathcal{P}_i} f_\pi = x_i, \quad \forall i \in [N];$$
(10a)

**Algorithm 1:** Greedy Algorithm - *GA*

1 Initialization: Let $S = \emptyset$, $\Pi = \emptyset$, $X = \{1, 2, \cdots, N\}$, $G' = G$
2 **while** *there exists driver $i \in X$ and path $\pi \in \cup_i \mathcal{P}_i$ from $s_i$ to $d_i$ with strictly positive profit $r_\pi > 0$* **do**
3    (a) Find the path $\pi^* = argmax_{\pi \in \cup_i \mathcal{P}_i} r_\pi$, such that $\pi^*$ has the maximum profit in the current graph $G'$. Let $\pi^*$ be the task list for driver $i^*$;
4    (b) Remove the source and destination nodes $(s_{i^*}, d_{i^*})$ of driver $i^*$ and all the task nodes in $\pi^*$ from the current graph $G'$;
5    (c) $S = S \cup i^*$, $\Pi = \Pi \cup \pi^*$, $X = X/i^*$;
6 **end**
7 Output the drivers in set $S$ and the selected paths (i.e. task lists) in $\Pi$.

$$\sum_{i=1}^{N} \sum_{\pi \in \mathcal{P}_i : m \in \pi} f_\pi \leq 1, \quad \forall m \in [M]; \tag{10b}$$

$$x_i \in \{0, 1\} \quad \forall i \in [N]; \tag{10c}$$

$$f_\pi \in \{0, 1\}, \quad \forall \pi \in \cup_i \mathcal{P}_i. \tag{10d}$$

Eq. (9) has the same interpretation as (4), as both calculate the drivers' total profits; (10a) means that each driver may choose 1 or 0 task list; (10b) means all the path chosen are node-disjoint that they do not intersect at any node.

### B. The Greedy Algorithm

To solve this problem, we first give an intuitive greedy algorithm, and then analyze the performance of the algorithm. Suppose all the paths chosen in the optimal solution have *strictly positive profit* as the paths with zero profit have no contribution to the drivers' total profits and therefore we will not chose them. In our greedy algorithm, we guarantee that the paths chosen also have *strictly positive profit*.

Note that for the step (a) in the iterations of Algorithm 1, we can find a path with highest-profit polynomially, since we can find the path with the highest value (or longest path) in a DAG within $O(N^2)$ if the values of the edges are all non-negative [18], where $N$ is the number of nodes in the DAG.

**Theorem 1.** *Algorithm 1 (i.e. GA) gives a feasible solution with $(\frac{1}{D+1})$-approximation ratio in polynomial time, where $D$ is the maximum number of nodes in a path (i.e. the diameter of the graph $G$). The ratio is tight.*

To prove Theorem 1, we first prove *GA* (i.e. Algorithm 1) can be finished in polynomial time. Then prove *GA* guarantee $(\frac{1}{D+1})$-approximation ratio, followed by an example which shows the tightness of this bound.

**Lemma 1.** *GA achieves a feasible solution of (4) within time complexity $O(N^2 M^2)$.*

*Proof.* To guarantee the feasibility, each driver has at most one task list. We can see that all the paths selected by *GA* have different source-destination pair. At the same time, each task node belongs to at most one selected path, and each selected path has *strictly positive profit*. To sum up, *GA* gives a feasible solution of (4).

*GA* terminates within at most $N$ iterations. Within each iteration, it costs $O(M^2)$ to find the highest-profit path for one driver by dynamic programming. Therefore it costs $O(NM^2)$ to find the highest-profit path in $G'$. So the total time complexity is no greater than $O(N^2 M^2)$. $\square$

**Lemma 2.** *GA guarantees an approximation ratio of $(\frac{1}{D+1})$.*

*Proof.* Suppose $\mathcal{B}$ is the set of paths selected by *GA* and $\mathcal{O}$ is the paths selected by the optimal solution (i.e. *OPT*). All the paths selected in $\mathcal{B}$ and $\mathcal{O}$ has strictly positive profit. We need to prove the following inequality that:

$$\sum_{\pi \in \mathcal{O}} r_\pi \leq (D+1) \cdot \sum_{\pi \in \mathcal{B}} r_\pi \tag{11}$$

Suppose *GA* terminates in $K$ iterations, $\{\pi_k\}_{k=1,2,\cdots,K}$ is the path selected by *GA* during the $k$-th iteration. We also define $G_\mathcal{B}^{k-1}$ as the updated graph just before the $k$-th iteration of *GA* (i.e. just after the $(k-1)$-th iteration of *GA*). Particularly, we have $G_\mathcal{B}^0 = G$.

In the following proof, we use "*intersect*" to express the relationship that two paths of either share the same node or share same the source-destination pair (i.e. belong to the same driver).

**Proposition 1.** *Every path in $\mathcal{O}$ must intersect with at least one path in $\mathcal{B}$.*

To prove this, if there exist a path $\pi'$ in $\mathcal{O}$ with positive profit, and $\pi'$ does not intersect with any paths in $\mathcal{B}$. Before *GA* terminates, $\pi'$ must be chosen by *GA*, otherwise we can insert an iteration of choosing $\pi'$ into *GA*. Therefore *GA* may have one more iteration and get a better result (i.e. by adding one more path with positive profit). Hence we can see that a contradiction is made here, such that Proposition 1 holds.

**Proposition 2.** *Every path in $\mathcal{B}$ intersects with at most $(D+1)$ paths in $\mathcal{O}$.*

Each path $\pi$ in $\mathcal{B}$ has at most $D$ (internal) nodes, intersect with at most 1 path in $\mathcal{O}$ at each node. So there are at most $D$ paths in $\mathcal{O}$ intersect with $\pi$ at its nodes. We can also see that at most 1 path share the same source-destination pair with $\pi$. Thus there are totally no more than $(D + 1)$ paths in $\mathcal{O}$ intersect with $\pi$, so Proposition 2 holds.

To further illustrate the relationship between the paths in $\mathcal{B}$ and $\mathcal{O}$. We use $\mathcal{O}_k$ to denote the set of paths in $\mathcal{O}$ that intersect with $\pi_k$. We also use notation $\bar{\mathcal{O}}_k$ to denote the set of paths in $\mathcal{O} \setminus \cup_{i=1}^k \mathcal{O}_k$, and use $G_\mathcal{O}^k$ to denote the updated graph after removing the nodes in $\cup_{i=1}^k \mathcal{O}_k$ from $G$. Start with $\mathcal{O}_0 = \emptyset$, $\bar{\mathcal{O}}_0 = \mathcal{O}$ and $G_\mathcal{O}^0 = G$. We apply a $K$-iteration process to construct $\mathcal{O}_k$, $\bar{\mathcal{O}}_k$ and $G_\mathcal{O}^k$, the process is shown in Algorithm 2.

**Proposition 3.**

$$\mathcal{O} = \cup_{k=1}^{K} \mathcal{O}_k \tag{12}$$

**Algorithm 2:** Construction of $\mathcal{O}_k$, $\bar{\mathcal{O}}_k$ and $G_{\mathcal{O}}^k$

1 Initialization: Let $\mathcal{O}_0 = \emptyset$, $\bar{\mathcal{O}}_0 = \mathcal{O}$ and $G_{\mathcal{O}}^0 = G$;
2 **for** *k = 1 to K* **do**
3    (a) Choose all the paths in $\bar{\mathcal{O}}_{k-1}$ which intersect with $\pi_k$, put these paths into $\mathcal{O}_k$;
4    (b) $\bar{\mathcal{O}}_k = \bar{\mathcal{O}}_{k-1} \setminus \mathcal{O}_k$;
5    (c) Remove all the nodes and source-destination pairs in $\pi_k$ and $\mathcal{O}_k$ from $G_{\mathcal{O}}^{k-1}$ to get the graph $G_{\mathcal{O}}^k$;
6    (d) Output $\mathcal{O}_k$, $\bar{\mathcal{O}}_k$ and $G_{\mathcal{O}}^k$;
7 **end**

Easy to find that $\mathcal{O}_k$ do not share any common nodes (i.e. task nodes or source/destination nodes) by the definition of $\mathcal{O}_k$. Using Proposition 1, we have $\bar{\mathcal{O}}_K = \emptyset$, otherwise the path in $\bar{\mathcal{O}}_K$ do not intersect with any paths in $\mathcal{B}$ which makes a contradiction with Proposition 1. Hence Proposition 3 holds.

**Proposition 4.**
$$\sum_{\pi \in \mathcal{O}_k} r_\pi \leq (D+1) \cdot r_{\pi_k}, \quad \forall k = 1, 2, \cdots, K \quad (13)$$

By the definition of $G_{\mathcal{O}}^k$ in Algorithm 2 we have $G_{\mathcal{O}}^k \subseteq G_{\mathcal{B}}^k$

$$G_{\mathcal{O}}^k \subseteq G_{\mathcal{B}}^k, \quad \forall k = 0, 1, 2, \cdots, K$$

Note that during the $k$-th iteration of *GA*, $\pi_k$ is the highest-profit path in $G_{\mathcal{B}}^{k-1}$, the profit of $\pi_k$ is greater than or equal to any paths in $G_{\mathcal{O}}^{k-1}$. Therefore

$$r_{\pi_k} \geq r_\pi, \forall \pi \in \mathcal{O}_k, \forall k = 1, 2, \cdots, K$$

By Proposition 2, $|\mathcal{O}_k| \leq D+1$, hence Proposition 4 holds. Finally, By Proposition 1 - Proposition 4, we have:

$$\sum_{\pi \in \mathcal{O}} r_\pi = \sum_{k=1}^{K} \sum_{\pi \in \mathcal{O}_k} r_\pi \leq \sum_{k=1}^{K} (D+1) \cdot r_{\pi_k} = (D+1) \cdot \sum_{\pi \in \mathcal{B}} r_\pi$$

Hence inequality (11) holds, the proof of Lemma 2 is done. Therefore $(\frac{1}{D+1})$ is an lower bound of the approximation ratio of *GA*. □

**Lemma 3.** $(\frac{1}{D+1})$ *is also the upper bound to the approximation ratio of* GA.

*Proof.* We construct an example, and the graph is shown in Figure 2. In the graph, we can see that there are totally $D+1$ tasks (i.e. the red nodes) and $D+1$ drivers with source-destination pairs $(s_1, d_1), (s_2, d_2), \cdots, (s_{D+1}, d_{D+1})$. Taking each task on the black path will gain $\frac{1}{D}$ profit for driver 1, but will gain $1-\epsilon$ profit for other drivers (i.e. the drivers in the set $\{2, 3, \cdots, D+1\}$). The graph has diameter exactly equals to $D$ with driver 1's longest path (i.e. the black path with $D$ tasks). From the construction in Algorithm 1, *GA* will only choose the longest path of driver 1 (i.e. the black path) with profit 1, and other drivers have no tasks to take. Therefore the drivers' total profits from *GA* is 1. On the other hand, the optimal solution will choose $(D+1)$ paths for the $D+1$ drivers, and each driver will take only one task and gain profit $1-\epsilon$. Therefore the optimal solution of the drivers' total profits is $(D+1)(1-\epsilon)$, and the approximation ratio in this example $\frac{1}{(D+1)(1-\epsilon)}$. Since $\epsilon$ can be infinitely small, the approximation ratio can be as bad as $(\frac{1}{1+D})$.

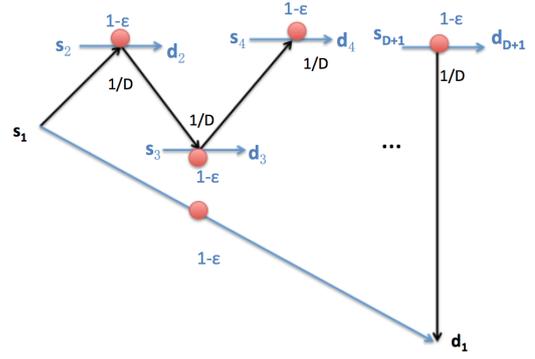

Fig. 2. is an example shows that *GA* performs no better than $(\frac{1}{D+1})$ theoretically.

Therefore *GA* performs no better than $(\frac{1}{D+1})$ theoretically, which shows $(\frac{1}{D+1})$ is also the upper bound to the approximation ratio of *GA*. □

Finally, by the proof of Lemma 1 to Lemma 3, we successfully prove Theorem 1.

### C. Discussion on our Approximation Algorithm

As an important problem in combinatorial optimization and graph theory, MDP problems attract substantial efforts on approximation algorithm design. Existing efforts on MDP problems most focus on the EDP (i.e. edge-disjoint paths) model, whereas our problem uses a specific *weighted node-disjoint paths* model. To the best of our knowledge, the state-of-the-art theoretical approximation ratio of EDP is given in [3]. The approximation ratio is $O\left(min(n^{2/3}, \sqrt{m})\right)$ for undirected graphs and $O\left(min(n^{4/5}, \sqrt{m})\right)$ for directed graph.

In the two-sided ride-sharing markets, we will not use the bounds in the existing work. Since both $n$ and $m$ are very large numbers in the ride-sharing markets: The value of $n$ is greater than the number of drivers, and the value of $m$ is greater than the number of tasks. Therefore we use specific approximation ratio of $(\frac{1}{D+1})$ for the ride-sharing markets. In our problem settings, $D$ (i.e. the diameter of the graph) represents the maximum number of possible tasks taken by a single driver during one working period. In the ride-sharing market, the value of $D$ is small, so Algorithm 1 can have fairly good performance. For instance, in the Uber market, a driver can only deal with a limited number of tasks during one working period [2] (i.e. 4 hours per working period on average). Furthermore, our algorithm works better for the Google's Waze Rider market, since Google is limiting drivers to two tasks a day, to and from work, and makes the price cheap, restricting drivers from making a living on the application [19]. Therefore in this market, we have $D = 1$ for one source-destination pair (i.e. from home to work or from work to home) of each driver, and our algorithm guarantees a $\frac{1}{2}$ approximation ratio for Google's Waze Rider market.

## V. ONLINE HEURISTIC SOLUTIONS

In some cases, especially for the Uber market, the tasks from the customers arrive in real-time. The platform and the drivers do not know the time or any other detailed information about a task in advance. Furthermore, the platform must give fast responses based on the real-time snapshots of the drivers and the tasks with very short response time.

The online problem is more challenging to solve. Our offline approximate algorithm is not applicable to the online scenarios since we do not have the information of all tasks in advance. Now we propose two heuristic algorithms that are applicable to both online or offline scenarios.

### A. Nearest drivers Heuristic

Once a task appears in the market, one of the most intuitive ways is to arrange this task to the driver who arrives at the source of the task *first*. From this point of view, the platform is prone to choose one of the drivers who may arrive at the source of the task as fast as possible. We estimate the fastest potential arrival time of each driver using the estimated distance divided by the average speed of the driver. We show the detailed algorithm in Algorithm 3. In Algorithm 3, the platform deals with the tasks in terms of their arrival time one by one, our algorithm keeps updating the status and location of the drivers. We can see from Algorithm 3 that if a driver finishes the task $m$ before the estimated finish time $\bar{t}_m^+$, she can drive to the source of her next task if the platform assigns a new task for her.

### B. Maximum Marginal Value Heuristic

Another heuristic algorithm is based on the marginal value added to the driver when a new task arrives and the driver takes it. This algorithm is similar to Algorithm 3. Each time a new task arrives, we choose an available candidate driver from the candidate set. The difference is the criterion of how to select the candidate from the set. To illustrate the criterion, we introduce a variable $\delta_{n,m}$, which is the marginal value added to the driver $n$ of task $m$, suppose the last task of driver $n$ is $m'$, and $m' = 0$ means $n$ has not taken any tasks yet.

$$\delta_{n,m} = p_m - (c_{n,m,-1} + \hat{c}_{n,m} + c_{n,m',m} - c_{n,m',-1}). \tag{14}$$

Each time we choose a driver from the candidate with maximum $\delta_{n,m}$. We show the algorithm in Algorithm 4. The structure of Algorithm 4 is the same as that of Algorithm 3.

Note that the maximum marginal value heuristic may also have another offline version that we can first sort the values of all the tasks if we know the entire data of all the task in advance with an offline setting. It will be more efficient to deal with the tasks which have higher values firstly.

---

**Algorithm 3:** Nearest Driver Heuristic - *Nearest*

1. Initialization: Unlock all the drivers, set their last tasks to 0, and let the candidate set be empty.
2. **while** *task $m$ arrives at time $\bar{t}_m$* **do**
3.    (a) Add the unlocked driver who can travel from their location to $\bar{s}_m$ during time $\bar{t}_m$ to $\bar{t}_m^-$ into the candidate set. Add the locked drivers who can travel from their current destination $\bar{d}_{m'}$ to $\bar{s}_m$ during time $\bar{t}_{m'}^+$ to $\bar{t}_m^-$ into the candidate set.
4.    (b) Choose the candidate driver $n^*$ in the candidate set who will arrive fastest to $\bar{s}_m$, if multiple, choose a random one. If the candidate set is empty, reject task $m$.
5.    (c) Set $n^*$ to lock status, set the last task of $n^*$ to $m$, update the location of driver $n^*$, reset candidate set to empty.
6.    (d) Task $m$ is done, the platform sends response to the customer and deals with the next task.
7. **end**

---

**Algorithm 4:** Maximum Marginal Value Heuristic - *max-Margin*

1. Initialization: Unlock all the drivers, set their last task to 0, and let the candidate set be empty.
2. **while** *task $m$ arrives at time $\bar{t}_m$* **do**
3.    (a) Add the unlocked drivers who can travel from its last destination to $\bar{s}_m$ during time $\bar{t}_m$ to $\bar{t}_m^-$ into the candidate set. Add the locked drivers who can travel from their last destination $\bar{d}_{m'}$ to $\bar{s}_m$ during time $\bar{t}_{m'}^+$ to $\bar{t}_m^-$ into the candidate set.
4.    (b) Choose the candidate driver $n^*$ in the candidate set such that $n^* = argmax(\delta_{n,m})$. If the candidate set is empty, reject task $m$.
5.    (c) Set $n^*$ to lock status, set the last task of $n^*$ to $m$, update the location of driver $n^*$, reset candidate set to empty.
6.    (d) Task $m$ is done, the platform sends response to the customer and deals with the next task.
7. **end**

---

## VI. PERFORMANCE EVALUATION

In this section, we evaluate the performance of our offline and online algorithms through real-trace simulations.

### A. Experiment setup

We use the dataset from ECML/PKDD 15 [20] including a complete year (from 01/07/2013 to 30/06/2014) of the trajectories for all the 442 taxis running in the city of Porto, Portugal. In the dataset, there are more than one million trip records with detailed information, including the timestamp of starting time and finishing time for each trip, polyline of the trip trajectory, and the driver ID. For the driver data, we can get the working time of each driver from her driver ID and the timestamps of her trips. We generate the source and destination of each driver using Monte Carlo method [21]. A special case that the driver has the same source and destination is meaningful, and it means that a driver leaves from a fixed place (may be her home) and returns after her daily work. This is referred to as the "home-work-home" model (i.e. the working model for full-time drivers on Uber) in the following results. There are also cases when the driver has different source and destination (e.g. the working model for part-time drivers on Google's Waze Rider), and we refer this working model as the "hitchhiking" model.

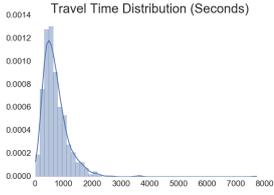

Fig. 3. Travel Time Distribution

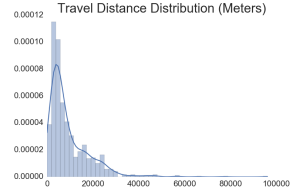

Fig. 4. Travel Distance Distribution

We clean the large dataset use the Pandas framework in Python [22]. For the trip data records, the traveling time distribution is shown in Fig 3 and the traveling distance distribution is shown in Fig. 4. We can find that the traveling time and traveling distance of the trips both exhibits the shape following the *power law distribution* [23].

The cost of each trip can be estimated by multiplying the total distance of the trip polyline and the unit price of gasoline (used as a constant). For the payoff of each trip to the driver, we use a simplified surge pricing model, that the payoff of task $m \in [M]$ is a linear equation of the travel distance and travel time multiplied by the *surge multiplier* $\alpha_m$, where $\alpha_m$ is dynamic changed based on real market scenarios. The simplified surge pricing is calculated by Eq. (15), in which $\beta_1$ and $\beta_2$ are both global constants.

$$p_m = \alpha_m \cdot (\beta_1 \cdot dis(\bar{s}_m, \bar{d}_m) + \beta_2 \cdot (\bar{t}_m^+ - \bar{t}_m^-)). \quad (15)$$

### B. Performance Ratio of Online/ Offline Algorithms

We use the drivers' total profits as the objective function of our algorithm. We use the offline relaxation results from $Z_f^*$ (defined in Section III) as the theoretical upper bound for our online and offline optimization problems. The performance ratio is $Z_f^*$ divided by the drivers' total profits achieved by the algorithms we design. For the evaluation of small-scale problems (e.g. for $n \leq 50$ and $m \leq 100$), we can use the integer programming solvers of CPLEX [24] or MOSEK [25] Optimizer to calculate the exact value of the best integer solution $Z^*$, and then use $Z^*$ as the upper bound for our optimization problems.

We select 1000 records during one day in the dataset. By gradually increasing the number of drivers available in the market from 20 to 300, we calculate the performance ratio of our algorithms. The results are shown in Fig. 5. We can find that our offline deterministic algorithm (Algorithm 1) has the best performance.s. For our online algorithms, the *maxMargin* algorithm (Algorithm 4) shows better result than that of the *Nearest* algorithm (Algorithm 3). Therefore the *maxMargin* algorithm is a good candidate for the online cases.

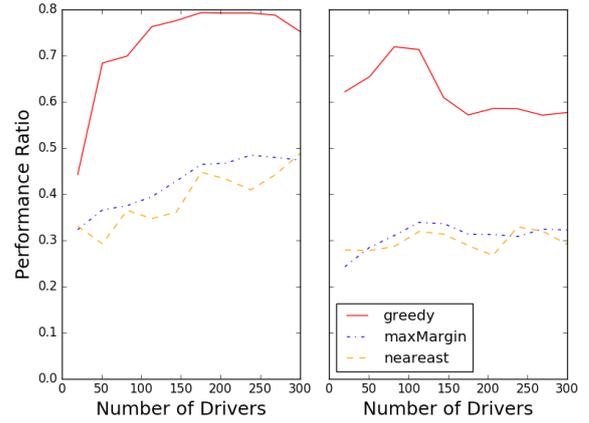

Fig. 5. The left figure shows the performance ratio of the "hitchhiking" model and the right figure shows the performance ratio of the "home-work-home" model. The four curves *Greedy*, *maxMargin*, and *Nearest* correspond to the deterministic offline algorithms, online maximum marginal value algorithm and online nearest worker algorithm accordingly.

Revisit the results in Fig. 5, we can see that almost all our algorithms achieve better performance ratio in the "hitchhiking" model than that of the "home-task-home" model. This implies that the "hitchhiking" model has better economic efficiency in real-life scenarios, it is more efficient to match riders and drivers already headed in the same direction (i.e. the working model of Google's Waze Rider).

### C. More Insights on the Ride-sharing Markets

We further discuss more simulation results which show the effectiveness of our algorithms applied in the real market. We will analyze the performance with different metrics for Algorithm 1 (the red line), Algorithm 4 (the blue line) and Algorithm 3 (the orange line). We use the same the simulation data as before and only consider the case of the general "hitchhiking" model that the drivers have random sources and destinations.

One interesting result is that when more drivers come into the market, the market becomes denser. Therefore more tasks will be served and more revenue will be generated. Fig. 6 shows that as the number of drivers increases, the total revenue generated in the market increases. Fig. 7 shows that as the number of drivers increases, the probability of a pending order to be served also increases. On the other hand, as the market become denser, it will lead to more competition and therefore the market congestion problem arises. Fig. 8 shows that as the number of drivers increases, the average payoff received by each driver declines. Fig. 9 shows similar results, as the as the number of drivers increases, the average tasks served by each driver also decreases.

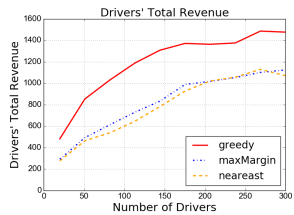

Fig. 6. Total revenue in the market

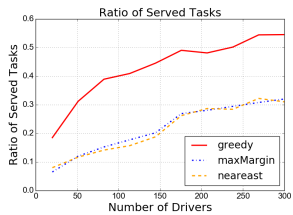

Fig. 7. Rate of served task

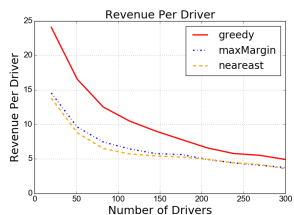

Fig. 8. Average revenue per worker

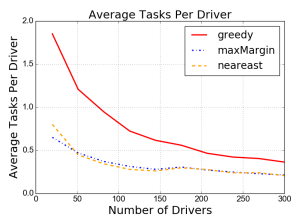

Fig. 9. Average tasks per worker

From the results above, we know that an effective matching market designer should make the market dense enough to ensure a high service rate, since a sparse market can not be efficient for most cases in the matching market. At the same time, the market designer should control the congestion level at a reasonable level, such that each participant in the market will be better off in the market. As we know, the Uber's *surge pricing mechanism* and the policy of Google's Waze Rider (i.e. limiting drivers and riders to two rides a day) are both promising ways to control the congestion in practice.

## VII. Concluding Remarks

Taxi and delivery markets are dynamic two-side markets that are highly associated with our daily life. Efficient algorithms are required to address the inefficient matching of the drivers and customers. To this end, we first propose a general framework to model the problem, and then provide a deterministic algorithm to solve the offline problem. This algorithm has a tight theoretical bound. For the online scenarios, we give two heuristic online algorithms. By real-trace simulations, we demonstrate the effectiveness to apply our algorithms in real markets. In the future, we hope to solve the online problem with non-heuristic algorithms and applied our algorithms in more models of two-sided markets.

## References

[1] J.-C. Rochet and J. Tirole, "Two-Sided Markets: An Overview," *mimeo: MIT*, 2004.
[2] M. K. Chen and M. Sheldon, "Dynamic Pricing in a Labor Market: Surge Pricing and Flexible Work on the Uber Platform," 2015.
[3] C. Chekuri and S. Khanna, "Edge disjoint paths revisited," in *Proceedings of the 14th Annual Symposium on Discrete Algorithms (SODA), pages 628–637.*, 2003.
[4] E. M. Azevedo1 and E. G. Weyl, "Matching markets in the digital age," *Science Vol. 352, Issue 6289, pp. 1056-1057*, 2016.
[5] L. Chen, A. Mislove, and C. Wilson, "Peeking Beneath the Hood of Uber," in *IMC '15: Proceedings of the 2015 ACM Conference on Internet Measurement Conference*, October 2015.
[6] J. Cramer and A. B. Krueger, "Disruptive Change in the Taxi Business: The Case of Uber," *American Economic Review, vol 106(5), pages 177-182.*, 2016.
[7] W. Z. andSatish V. Ukkusuri, "Optimal Fleet Size and Fare Setting in Emerging Taxi Markets with Stochastic Demand," *Computer-Aided Civil and Infrastructure Engineering*, 2016.
[8] M. Desrochers, J. Lenstra, M. Savelsbergh, and F. Soumis., "Vehicle routing with time windows: optimization and approximation," *B.L. Golden, A.A. Assad (eds.). Vehicle Routing: Methods and Studies, North-Holland, Amsterdam, pages 65–84*, 1999.
[9] N. Bansal, A. Blum, S. Chawla, and A. Meyerson, "Approximation Algorithms for Deadline-TSP and Vehicle Routing with Time-Windows," in *STOC '04: Proceedings of the thirty-sixth annual ACM symposium on Theory of computing*, 2004.
[10] M. Desrochers, J. Desrosiers, and M. Solomon., "A new optimization algorithm for the vehicle routing problem with time windows," *Operations Research, 40:342–354*, 1992.
[11] M. Savelsbergh, "Local search for routing problems with time windows," *Annals of Operations Research, 4:285–305*, 1985.
[12] K. Tan, L. Lee, and Q. Zhu., "Heuristic methods for vehicle routing problem with time windows," in *Artificial Intelligence in Engineering*, 2001.
[13] R. V. S. R. Thangiah, I. H. Osman and T. Sun., "Algorithms for the vehicle routing problems with time deadlines," *American Journal of Mathematical and Management Sciences 13:323–355*, 1993.
[14] A. Kolen, A. R. Kan, and H. Trienekens, "Vehicle routing with time windows," *Operations Research, 35:266–273*, 1987.
[15] V. Guruswami, S. Khanna, R. Rajaraman, B. Shepherd, and M. Yannakakis, "Near-Optimal hardness results and approximation algorithms for edge-disjoint paths and related problems," in *STOC '99: Proceedings of the thirty-first annual ACM symposium on Theory of computing*, 1999.
[16] S. Fortune, J. Hopcroft, and J. Wyllie, "The directed subgraph homeomorphism problem," *Theoretical Computer Science*, 1980.
[17] S. Boyd and L. Vandenberghe, *Convex Optimization*. Cambridge University Press, 2004.
[18] M. Khan, *Longest path in a directed acyclic graph*. CCE 221, Emory University, 2011.
[19] *Google Quietly Expands Ride-Sharing Service*. http://www.wsj.com/articles/googlequietlyexpandsridesharingservice1475179947.
[20] Kaggle, "ECML/PKDD 15 datasets in Kaggle," *https://www.kaggle.com/c/pkdd-15-predict-taxi-service-trajectory-i/*, 2015.
[21] R. Y. Rubinstein and D. P. Kroese, *Simulation and the Monte Carlo Method*. Wiley, 2007.
[22] Pandas, "Pandas – Python Data Analysis Library," *http://pandas.pydata.org*, 2016.
[23] D. Easley and J. Kleinberg, *Networks, Crowds, and Markets: Reasoning About a Highly Connected World*. Cambridge University Press, 2010.
[24] *CPLEX Optimizer*. https://www-01.ibm.com/software/commerce/optimization/cplex-optimizer.
[25] *MOSEK Optimizer*. https://www.mosek.com/.